\documentclass[%
 reprint,
groupedaddress,
nofootinbib,
amsmath,amssymb,
aps,
]{revtex4-2}

\usepackage{graphicx,xcolor}
\usepackage{dcolumn}
\usepackage{bm,bbm,xfrac,mathtools}
\usepackage{hyperref}

\usepackage{tabularx}

\let\Gamma\varGamma
\let\Delta\varDelta
\let\Theta\varTheta
\let\Xi\varXi
\let\Pi\varPi
\let\Sigma\varSigma
\let\Upsilon\varUpsilon
\let\Phi\varPhi
\let\Psi\varPsi
\let\Omega\varOmega

\newcommand{\de}{\mathrm{d}}

\newcommand{\om}{\Omega_{{\rm m},0}}
\newcommand{\bs}{\ensuremath{b\sigma_8}}
\newcommand{\fs}{\ensuremath{f\!\sigma_8}}
\renewcommand{\ss}[1][]{\ensuremath{\Omega\sigma_{8#1}}}
\renewcommand{\ap}{\ensuremath{\alpha_\perp}}
\newcommand{\lcdm}{\(\Lambda\)CDM}
\renewcommand{\vec}[1]{\ensuremath{\boldsymbol{#1}}}
\newcommand{\tens}[1]{\ensuremath{\mathsf{#1}}}

\begin{document}

\preprint{APS/123-QED}

\title{A novel method for unbiased measurements of growth with cosmic shear}

\author{Stefano Camera}
\email{stefano.camera@unito.it}
\affiliation{
Dipartimento di Fisica, Universit\`a degli Studi di Torino, 10125 Torino, Italy\\
INFN -- Istituto Nazionale di Fisica Nucleare, Sezione di Torino, 10125 Torino, Italy\\
INAF -- Istituto Nazionale di Astrofisica, Osservatorio Astrofisico di Torino, 10025 Pino Torinese, Italy\\
Department of Physics \& Astronomy, University of the Western Cape, Cape Town 7535, South Africa
}%



\begin{abstract}
I present a new technique for the measurement of the growth of cosmic structures via the power spectrum of weak lensing cosmic shear. It is based on a template-fitting approach, where a redshift-dependent amplitude of lensing modulates a fixed template power spectrum. Such an amplitude, which is promoted to a free parameter and fit against tomographic cosmic shear data, reads \(D(z)\,\sigma_8\,\om\eqqcolon\ss(z)\), with \(D(z)\) the growth factor, \(\sigma_8\) a proxy for the overall amplitude of the matter power spectrum, and \(\om\) the present-day matter abundance. I show that this method is able to correctly reconstruct \ss\ at the per cent level across redshift, thus allowing us to measure the growth of structures unbiased by observing discrete tracers. Moreover, I only makes use of measurements on linear scales. The method is highly complementary to measurements of the bias and growth, \bs\ and \fs, from galaxy clustering analysis. I also demonstrate that the method is robust against an incorrect choice of cosmological parameters in the template, thanks to the inclusion of an Alcock-Paczy\'nski parameter. 
\end{abstract}

\maketitle


\section{Introduction}
The last decades have witnessed the establishment of a standard model for cosmology, dubbed \lcdm\ after its main ingredients: a cosmological constant, \(\Lambda\), and cold dark matter. The former is necessary to explain the late-time accelerated expansion of the Universe, whilst the latter is a non-standard, exotic matter component which accounts for the observed properties of the cosmic large-scale structure \citep{2022arXiv220108666L}.

Despite the undisputable success of \lcdm\ in explaining a wide variety of cosmological observations---from the anisotropies in the cosmic microwave background radiation at early times to the clustering of collapsed objects like galaxies and clusters of galaxies---fundamental questions remain unanswered. Foremost amongst them is the nature of dark matter and whether the cosmological constant is really a vacuum energy, or rather reflects an evolving `dark energy' component. Moreover, the possibility that the effects we ascribe to either or both these dark components are in fact due to an incorrect knowledge of how gravity works on cosmological scales, is still open \citep[for a review, see][]{2016PDU....12...56B}.

To tackle the aforementioned issues, the cosmic large-scale structure is a perfect arena of study. Two main observables have been regarded as the most promising in this respect: galaxy clustering and cosmic shear. The former employs galaxies as discrete tracers of the underlying matter distribution to track its growth across cosmic time.\footnote{Albeit galaxies are by far the most widely used, other compelling possibilities exist, for instance galaxy clusters, galaxy peculiar velocities, or neutral hydrogen intensity mapping.} The latter exploits correlations in the observed shapes of galaxies caused by (weak) gravitational lensing due to the intervening large-scale structure, to probe both the spacetime geometry and the cosmic gravitational potential. Besides being powerful in themselves, such two probes are highly complementary to each other, reason for which a strong experimental effort is currently ongoing to pursue commensal observational campaigns \citep[see e.g.][]{Blanchard:2019oqi,2019ApJS..242....2C,2020PASA...37....7S}.

To the aim of extracting the most information from joint observations, it is clearly paramount to be able to perform combined data analyses in a self-consistent way. However, this historically represented a problem, for the two observables are treated in different ways. First is the fact that clustering is inherently three-dimensional, with galaxy positions being correlated in configuration or Fourier space. Oppositely, shear is an integrated effect, with galaxy ellipticities being correlated on the celestial sphere or in harmonic space. Secondly, in galaxy clustering it has been customary to fit the observed correlation function/power spectrum in terms of two redshift-dependent amplitudes, \(b\sigma_8(z)\coloneqq b(z)\,D(z)\,\sigma_8\) and \(f\!\sigma_8(z)\coloneqq f(z)\,D(z)\,\sigma_8\), where \(\sigma_8\) is the rms of matter fluctuations on spheres of \(8\,\mathrm{Mpc}\,h^{-1}\) radius, \(b(z)\) is the linear galaxy bias, and \(f(z)\coloneqq-\de\ln D/\de\ln(1+z)\) is the growth rate, with \(D(z)\) being the growth factor. Then, constraints on cosmological parameters are extracted from \bs\ and \fs\ (plus the so-called Alcock-Paczy\'nski parameters). On the other hand, in cosmic shear the observed correlation function/power spectrum is directly fit against a theoretical prediction as a function of cosmological parameters.

Whilst it is not possible to recast cosmic shear measurements in three dimensions, studies of galaxy clustering in e.g. harmonic space have been attempted several times. The price to pay, to make the problem computationally tractable, is however a severe loss of constraining power due to the necessity to collapse the radial information into redshift bins. To circumvent this problem, \citet{2021arXiv210700026T} proposed a novel \bs-\fs\ template-fit approach for the galaxy clustering harmonic-space power spectrum. This allows for a finer slicing in redshift of the galaxy distribution at a very little computational cost. Here, I extend this formalism to cosmic shear, setting the scene for wholly self-consistent data analyses.

Throughout the paper, I adopt a fiducial flat $\Lambda$CDM cosmology with: total matter abundance at present, \(\om=0.3\); Hubble constant, \(h\coloneqq H_0/(100\,\mathrm{km\,s^{-1}\,Mpc^{-1}})=0.67\); amplitude and tilt of the primordial power spectrum, \(A_{\rm s}=2\times10^{-9}\) and \(n_{\rm s}=0.96\); and total neutrino mass, \(M_\nu=0.06\,\mathrm{eV}\). Moreover, I employ the following index convention:
\begin{itemize}
    \item Lower-case, Latin letters \(i,j=1\ldots N_z\) denote redshift bins;
    \item \(\ell=1\ldots N_\ell\) and \(m=-\ell\ldots\ell\) (and variants) denote harmonic expansion multipoles;
    \item Upper-case, Latin letters \(A,B=1\ldots N_{\rm d}\) label data points and, consequently, theoretical predictions corresponding to them;
    \item Lower-case, Greek letters \(\mu,\nu=1\ldots N_\vartheta\) label the parameters of the model.
\end{itemize}

Spatial vectors, like three-dimensional positions or wave-vectors, will be rendered in boldface, e.g.\ \(\vec r=r\,\hat{\vec r}\), with magnitude \(r=|\vec r|\) and direction \(\hat{\vec r}\). Either than that, to avoid confusion I shall use matrix notation only with respect to tomographic bins; for instance, \(\vec{x}=\{x_i\}\) and \(\tens{Y}=\{Y_{ij}\}\) are an array and a matrix of bin-dependent quantities, respectively. Any other non-scalar quantity will instead be rendered only as a collection of its elements, say \(\{\zeta_\alpha\}\) or \(\{Z_{AB}\}\). Einstein convention for summation over equal indexes is assumed unless otherwise stated.

\section{Cosmic shear}
Cosmic shear, \(\gamma\), is one of the two effects sourced by scalar cosmological perturbations which weak lensing distortions can be decomposed into, alongside convergence, \(\kappa\) \citep{2006asup.book..213B}. They are both related to the lensing potential \(\phi\), that is the projection of the Weyl potential \(\Upsilon\), along the line-of-sight direction, \(\hat{\vec r}\). Such a projection is a weighted integral, namely
\begin{equation}
    \phi(\hat{\vec r})\coloneqq\int\de r\,q(r)\,\Upsilon(r\,\hat{\vec r},r)\;,\label{eq:phi}
\end{equation}
where the weight \(q\) is usually referred to as lensing efficiency.\footnote{In \autoref{eq:phi}, the former argument of \(\Upsilon\) is the spatial coordinate \(\vec r=r\,\hat{\vec r}\), whereas the latter is the time coordinate, for which I use the radial comoving distance \(r\) as a proxy, given that the conformal time \(\eta\) is such that \(r=\eta_0-\eta\), with \(\eta_0\) being conformal time at present.} The lensing efficiency \citep[see also][]{2020OJAp....3E...6T} itself depends upon \(n(z)\), the distribution in redshift of the sources whose images are being distorted by the intervening cosmic large-scale structure, and reads
\begin{equation}
    q(r)=\int_r^\infty\de x\,\frac{x-r}{x}\,n(x)\;,
\end{equation}
where \(n(z)\) has been normalised to unit area. Notice that \(n(z)\,\de z=n(r)\,\de r\), with \(\de r/\de z=1/H\), \(H(z)\) being the Hubble factor at redshift \(z\).

If we denote by \(\bar{\vec\theta}\) the unlensed coordinate system at source, and by \(\vec\theta\) the lensed coordinates at observer, at first order we have
\begin{equation}
    \bar{\vec\theta}=\tens A\,\vec\theta\;,
\end{equation}
where
\begin{equation}
    \tens A=\left[
    \begin{array}{cc}
        1-\kappa-\Re(\gamma) & \Im(\gamma) \\
         -\Im(\gamma) & 1-\kappa+\Re(\gamma)
    \end{array}
    \right]
\end{equation}
is the distortion matrix, \(\Re(x)\) and \(\Im(x)\) respectively denoting the real and imaginary parts of a complex quantity \(x\). It is clear then that convergence, responsible for an overall amplification/shrinking in size of the image, is a scalar quantity, whereas shear is a complex pseudo-vector (or spin-2) and its real and imaginary parts distort images by squeezing them respectively along the coordinate axes and with angle of \(45^\circ\) with respect to them. Since the distortion matrix is directly related to (derivatives on the image plane of) the lensing potential, we can formally write \(\kappa=\eth\eth^\ast\phi/2\) and \(\gamma=\eth^2\phi/2\), where \(\eth\) is the spin-raising operator, which augments the spin of a quantity by \(+1\) each time it is applied, or diminishes it by the same amount if its complex conjugate is used instead \citep{1967JMP.....8.2155G}.

To obtain the shear power spectrum we first need to expand the shear into spin-weighted spherical harmonics \citep[][]{2005PhRvD..72b3516C,2017JCAP...05..014L},
\begin{equation}
    \gamma(\hat{\vec r})=\sum_{\ell=0}^\infty\,\sum_{m=-\ell}^\ell\,\left(\epsilon_{\ell m}+{\rm i}\,\beta_{\ell m}\right)\,_2Y_{\ell m}(\hat{\vec r})\;,
\end{equation}
where \(\epsilon_{\ell m}\) and \(\beta_{\ell m}\) are respectively the expansion coefficients for shear \(E\)- and \(B\)-modes, and \(_sY_{\ell m}\) is the spin-\(s\) spherical harmonic function. Then, if we have two shear maps averaged over different redshift shells labelled by indexes \(i\) and \(j\), we eventually get to the well-known expression for the tomographic shear \(E\)-mode power spectrum
\begin{align}
    S^{\epsilon\epsilon}_{ij,\ell}&\coloneqq\left\langle\epsilon_{i,\ell m}^{\phantom{ast}}\,\epsilon_{j,\ell m}^\ast\right\rangle\label{eq:Cl_DEF}\\
    &=\frac2\pi\int\de k\;k^2\,P_{\rm lin}(k)\,\epsilon_{i,\ell}(k)\,\epsilon_{j,\ell}(k)\;.\label{eq:Cl}
\end{align}
There, \(\epsilon_{i,\ell m}\) are the coefficients of the spin-weighted spherical harmonic expansion of shear \(E\)-modes in the \(i\)th redhisft bin, namely \(\epsilon_i(\hat{\vec r})\), \(P_{\rm lin}\) is today's linear matter power spectrum, and \(\epsilon_{i,\ell}(k)\) are the shear kernels. In the \lcdm\ model and in the linear regime of cosmological perturbations, they can be written as
\begin{align}
    \epsilon_{i,\ell}(k)&=\int\frac{\de z}{H(z)}\,D(z)\,q_i(z)\,\check{\jmath}_\ell[k,r(z)]\;,\label{eq:epsilon_lk}\\
    \check{\jmath}_\ell(k,r)&=-\frac32\,\om\, H_0^2\,\sqrt{\frac{(\ell+2)!}{(\ell-2)!}}\,\left[1+z(r)\right]\,\frac{j_\ell(k\,r)}{k^2\,r}\;,
\end{align}
with \(j_\ell\) being the spherical Bessel function of order \(\ell\).

\section{Aims and methodology}
\begin{figure*}
    \centering
    \includegraphics[width=\textwidth]{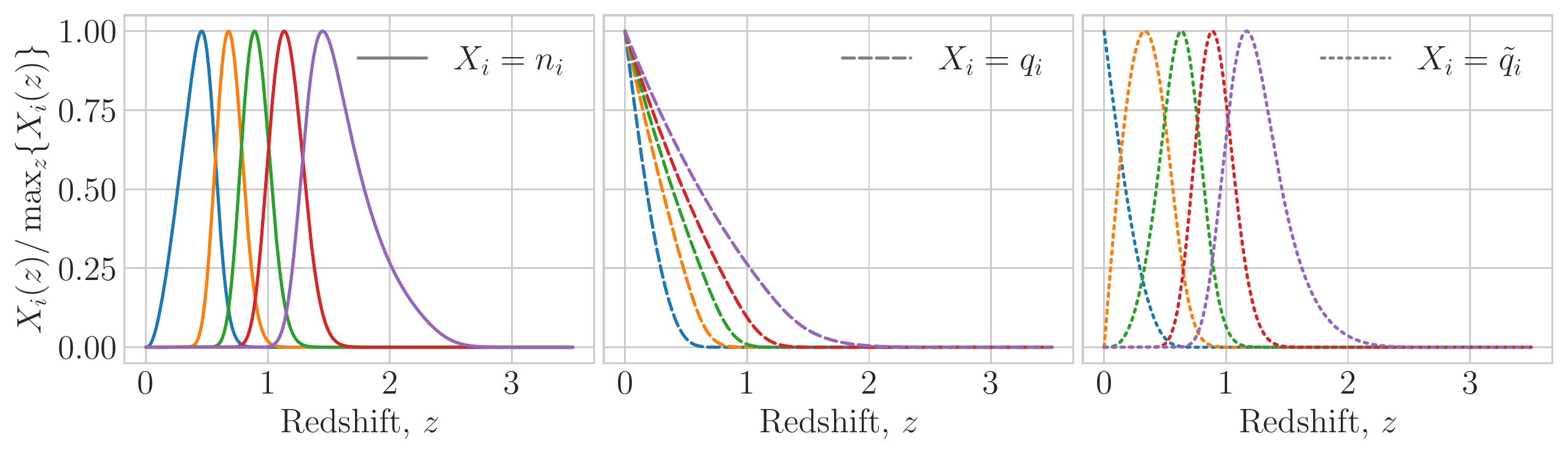}
    \caption{\textit{Leftmost panel:} binned redshift distribution of sources, \(n_i(z)\). \textit{Central panel:} corresponding lensing efficiency, \(q_i(z)\). \textit{Rightmost panel:} BNT-transformed lensing efficiencies, \(\tilde q_i(z)\). All curves are normalised to their peak, to facilitate comparison.}
    \label{fig:Xz}
\end{figure*}
The present paper builds upon the template-fitting method proposed by \citet{2021arXiv210700026T} in the context of measurements of galaxy clustering with harmonic-space power spectra.

Here, I extend for the first time this template fitting to cosmic shear power spectra. The advantage is that lensing is an unbiased tracer of the (projected) matter distribution. Moreover, as it seems clear by looking at \autoref{eq:epsilon_lk}, shear is directly sensitive to \(D(z)\,\sigma_8\,\om\,h^2\). However, as shown by \citet{2021MNRAS.505.4935H}, the dependency upon the Hubble constant is only apparent, as in the linear regime it cancels out exactly because of the \(h\)-dependence of the various quantities inside the integral in \autoref{eq:Cl}. Therefore, the actual proportionality is captured by the following redshift-dependent quantity:
\begin{equation}
    \ss(z)\coloneqq D(z)\,\sigma_8\,\om\;,
\end{equation}
akin to \(\bs(z)\) and \(\fs(z)\). Hence, I can rewrite \autoref{eq:Cl} as (no summation over equal indexes)
\begin{equation}
    S^{\epsilon\epsilon}_{ij,\ell}\simeq
    \Omega\sigma_{8,i}\,\ss[,j]\,T^{\epsilon\epsilon}_{ij,\ell}\;,\label{eq:Cl_ours}
\end{equation}
where \(\ss[,i]\equiv\ss(\bar z_i)\) and I have defined the template
\begin{multline}
    T^{\epsilon\epsilon}_{ij,\ell}\coloneqq\frac2\pi\int\de k\,\de r_1\,\de r_2\\\times k^2\,\frac{P_{\rm lin}(k)}{\om^2\,\sigma_8^2}\,q_i(r_1)\,q_j(r_2)\,\check{\jmath}_\ell(k,r_1)\,\check{\jmath}_\ell(k,r_2)\;.\label{eq:Cl_phiphi}
\end{multline}
Above, \(\bar z_i\) is the effective redshift of each bin, computed via
\begin{equation}
    \bar z_i=\frac{\int\de z\,n_i(z)\,z}{\int\de z\,n_i(z)}\;.
\end{equation}

Compared to galaxy clustering, in the case of shear there is an additional complication, namely the fact that lensing kernels are broad in redshift, for lensing is an integrated quantity. This represents a problem, in that the for the template to reproduce the actual power spectrum at the desired accuracy, the redshift-dependent amplitude \(\ss(z)\) should vary slowly within the redshift bin, so that it can be taken out of the integral in \autoref{eq:Cl} and become amplitudes parameters as in \autoref{eq:Cl_ours}. To appreciate this issue, consider a population of sources described e.g.\ by the following functional form \citep{1994MNRAS.270..245S},
\begin{equation}
    n(z)\propto\left(\frac{z}{z_0}\right)^2\,\exp\left[-\left(\frac{z}{z_0}\right)^{3/2}\right]\;,
\end{equation}
where I take \(z_0=0.9/\sqrt{2}\) as expected for next-generation cosmic shear surveys \citep{Blanchard:2019oqi,2019ApJS..242....2C}.  Then, split this distribution into, say, \(N_z=5\) equi-populated redshift bins, convolving the bins with a reasonable (Gaussian) photometric redshift error of \(0.05\,(1+z)\). In \autoref{fig:Xz}, I show the binned galaxy redshift distribution, \(n_i(z)\) (leftmost panel, solid curves), alongside its corresponding lensing efficiencies, \(q_i(z)\) (central panel, dashed curves). As it is clear, even for bins that are far apart, like the first and the last, the lensing signal is going to be highly correlated, due to the inherent shape of the efficiency function.

Compared to galaxy clustering, this represents an additional complication, in that for the template to reproduce the actual power spectrum at the desired accuracy, the redshift-dependent amplitude \(\ss(z)\) should vary slowly within the redshift bin. Thus, it can be taken out of the integral in \autoref{eq:Cl} as \(\ss[,i]\). To overcome this issue, I take advantage of the nulling technique proposed by \citet*{2014MNRAS.445.1526B} to localise the lensing kernels \citep[see also][]{2021PhRvD.103d3531T,2018PhRvD..98h3514T}. To overcome this issue, I take advantage of the nulling technique proposed by \citet*{2014MNRAS.445.1526B} to localise the lensing kernels \citep[see also][]{2021PhRvD.103d3531T,2018PhRvD..98h3514T}. This is performed via the so-called BNT matrix \(\tens M\), such that the BNT-transformed lensing efficiency becomes \(\tilde{\vec q}(z)=\tens M\,\vec q(z)\). The result of the BNT transform can be appreciated by looking at the rightmost panel of \autoref{fig:Xz}, where the \(\tilde q_i(z)\)'s (dotted curves) clearly show the recovery of localisation.

Hence, I proceed as follows: \(i)\) I compute the template tomographic matrix \(\tens T^{\epsilon\epsilon}_\ell\) as in \autoref{eq:Cl_phiphi}; \(ii)\) I then multiply its \(i\)-\(j\) entry by \(\ss[,i]\,\ss[,j]\), thus obtaining \(\tens S^{\epsilon\epsilon}_\ell\) as in \autoref{eq:Cl_ours}; \(iii)\) and finally, at each fixed \(\ell\) I apply the BNT transform as \(\tens M\,\tens S^{\epsilon\epsilon}_\ell\,\tens M^{\sf T}\).

One last remark has to be made before testing this template with synthetic data. So far, I have implicitly assumed that the cosmological information is contained within the redshift-dependent amplitude \(\ss(z)\). However, both the lensing efficiency and the power spectrum in \autoref{eq:Cl} do depend upon cosmological parameters. As a consequence, if the reference cosmology assumed to compute \autoref{eq:Cl_ours}-\ref{eq:Cl_phiphi} is not the correct one, I am likely to obtain incorrect results. A way out of this impasse is represented by the Alcock-Paczy\'nski parameter, which is customarily included in galaxy clustering analyses. This is due to the fact that one has to assume a reference cosmology to translate measured angular positions and redshifts into the physical positions of the galaxies.

In the present case of the harmonic-space power spectrum, we in fact only consider angles and redshifts. However, we know that a given physical size \(L\) subtended by an angle \(\theta\) corresponds to an angular diameter distance \(D_{\rm A}=L/\theta\). Therefore, if the reference cosmology is not the real one, we have to translate a measured feature at a given angular scale \(\ell\) into the one it would correspond to in the reference cosmology. Some little algebra shows that \(\ell^{\rm(ref)}=\ap^{-1}\,\ell\), with the usual definition
\begin{equation}
    \ap\coloneqq\frac{D_{\rm A}}{D_{\rm A}^{\rm(ref)}}\;.
\end{equation}
Moreover, we should remember that the harmonic-space power spectrum has units of steradians. Therefore, to relate the theoretical prediction in the reference cosmology to the real one, we have to transform the solid angle \(\theta^2=[\ap^{-2}\,\theta^{\rm(ref)}]^2\). Thence, I recast
\begin{equation}
    S^{\gamma\gamma}_{ij,\ell}\simeq\ap^{-2}\,\ss[,i]\,\ss[,j]\,T^{\gamma\gamma}_{ij}\left(\ap^{-1}\,\ell\right)\;,\label{eq:Cl_ours_AP}
\end{equation}
where I have now promoted \(\ell^{\rm(ref)}=\ap^{-1}\,\ell\) to a continuous variable, for which I interpolate the template of the tomographic power spectrum, \(\tens T^{\epsilon\epsilon}\).

\section{Validation and discussion of results}
To assess the performance of the method, I set up a mock-data analysis. The parameter set I fit for is \(\{\vartheta_\mu\}=\{\ss[,i]\}\cup\{\ap\}\). A synthetic data vector \(\{d_A\}=\{\tens M\,\tens S^{\epsilon\epsilon}_\ell\,\tens M^{\sf T}\}\), \(A=1\ldots N_{\rm d}\), is created by computing \(\tens S^{\epsilon\epsilon}_\ell\) exactly via \autoref{eq:Cl}. In turn, this is compared to the BNT-transformed template given by \autoref{eq:Cl_ours_AP}. I implement a Markov chain Monte Carlo sampling of the \(N_z+1\)-dimensional parameter space by minimising the chi-squared
\begin{equation}
    \chi^2\left(\{\vartheta_\mu\}\right)=\sum_{A,B=1}^{N_{\rm d}}\left[d_A - m_A\left(\{\vartheta_\mu\}\right)\right]\,\Gamma_{AB}\,\left[d_B - m_B\left(\{\vartheta_\mu\}\right)\right]\;.\label{eq:X2}
\end{equation}
Here, \(\{m_A\}\) are the model predictions, dependent upon the parameters \(\{\vartheta_\mu\}\), and \(\Gamma_{AB}\) denotes the elements of the precision matrix, i.e.\ the inverse of the data covariance matrix, \(\{\Sigma_{AB}\}\).


It is worth noticing that the factorisation of \ss\ relies on the separability of the \(k\) and \(z\) dependence of the matter power spectrum, which is strictly true only in the linear regime. Therefore, the analysis is limited to linear scales, \(k<k_{\rm nl}=0.2\,h\,\mathrm{Mpc}^{-1}\). To be conservative, I do not consider that \(k_{\rm nl}\) increases with redshift. Then, I estimate the maximum multipole from the Limber approximation, taking advantage of the fact that \(k\sim(\ell+1/2)/r(z)\) holds for \(\ell\gg1\). Hence, I define a set of redshift-dependent maximum multipoles, \(\ell_{\rm max}^i=\lfloor k_{\rm nl}\,r(\bar{z}_i)-1/2\rfloor\), where \(\lfloor x\rfloor\) is the floor function, that associates to a real number \(x\) the greatest integer less than or equal to \(x\). In the present case, they take values \(\{156,\,339,\,428,\,517,\,705\}\). Then, I adopt a conservative approach and for the analysis set the multipole range \(\ell=2\ldots156\). In other words, I take the smallest of the \(\ell_{\rm max}^i\)'s for all the redshift bins. Dependence upon the maximum multipole, as well as on the number of redshift bins, will be discussed later.

For simplicity, I assume that the covariance of the data be Gaussian. I consider the measured tomographic power spectrum to be the sum of the underlying signal plus some (statistical or instrumental) noise, viz.\
\begin{equation}
    \tens{C}^{\epsilon\epsilon}_\ell=\tens{S}^{\epsilon\epsilon}_\ell+\tens{N}^{\epsilon\epsilon}_\ell\;.
\end{equation}
The noise is approximated as Poissonian, rescaled by the variance of measured galaxies' ellipticities, \(\sigma_\epsilon^2\). Namely,
\begin{equation}
    N^{\epsilon\epsilon}_{ij,\ell}=\frac{\sigma_\epsilon^2}{\bar n_{\rm g}^i}\,\delta^{\rm(K)}_{ij}\;,
\end{equation}
where \(\bar n_{\rm g}^i\) is the average number density of galaxies per steradian in the \(i\)th redshift bin and \(\delta^{\rm(K)}\) is the Kronecker symbol. I take the common choice \(\sigma_\epsilon=0.3\) and assume \(\bar n_{\rm g}^i=30\,\mathrm{arcmin}^{-2}\). Thus, we can write
\begin{equation}
{\rm cov}\left(C^{\epsilon\epsilon}_{ij,\ell},C^{\epsilon\epsilon}_{ab,\ell^\prime}\right)=\frac{C^{\epsilon\epsilon}_{ia,\ell}\,C^{\epsilon\epsilon}_{jb,\ell}+C^{\epsilon\epsilon}_{ib,\ell}\,C^{\epsilon\epsilon}_{ja,\ell}}{2\,\ell+1}\,\delta^{\rm(K)}_{\ell\ell^\prime}\;.\label{eq:cov}
\end{equation}
After computing \autoref{eq:cov}, we stack the \((i,j)\) bin-pair index and the \(\ell\) multipole to construct the covariance matrix, whose entries are \(\Sigma_{AB}\). Finally, binning the synthetic data into multipole bins of width \(\varDelta\ell\) and accounting for partial sky coverage corresponds to the substitution \(\{\Sigma_{AB}\}\to\{\Sigma_{AB}\}/(\varDelta\ell\,f_{\rm sky})\). For simplicity, I consider \(f_{\rm sky}=1\). Also, I adopt \(\varDelta\ell=5\), i.e.\ a few times \(1/f_{\rm sky}\) as commonly done.

To test the performance of the method in different scenarios, I adopt in turn different reference cosmologies (on top of our fiducial cosmology), as in \autoref{tab:params}. I emphasise that I implement the BNT transform in all cases but Case 0, which is used for comparison.
\begin{table}
    \centering
    \caption{Values of cosmological parameters assumed as reference in the various cases considered. Case 0 and 1 are in fact the same, but for no BNT transform being applied to the former.}
    \begin{tabularx}{\columnwidth}{lXXXXX}
        Case & 0 & 1 & 2 & 3 & 4 \\
        \hline\hline
        \(h^{\rm(ref)}/h\) & 1 & 1 & 1 & 1.2 & 1.2 \\ 
        \hline
        \(\om^{\rm(ref)}/\om\) & 1 & 1 & 0.8 & 1 & 0.8
    \end{tabularx}
    \label{tab:params}
\end{table}

\begin{figure}
    \centering
    \includegraphics[width=\columnwidth]{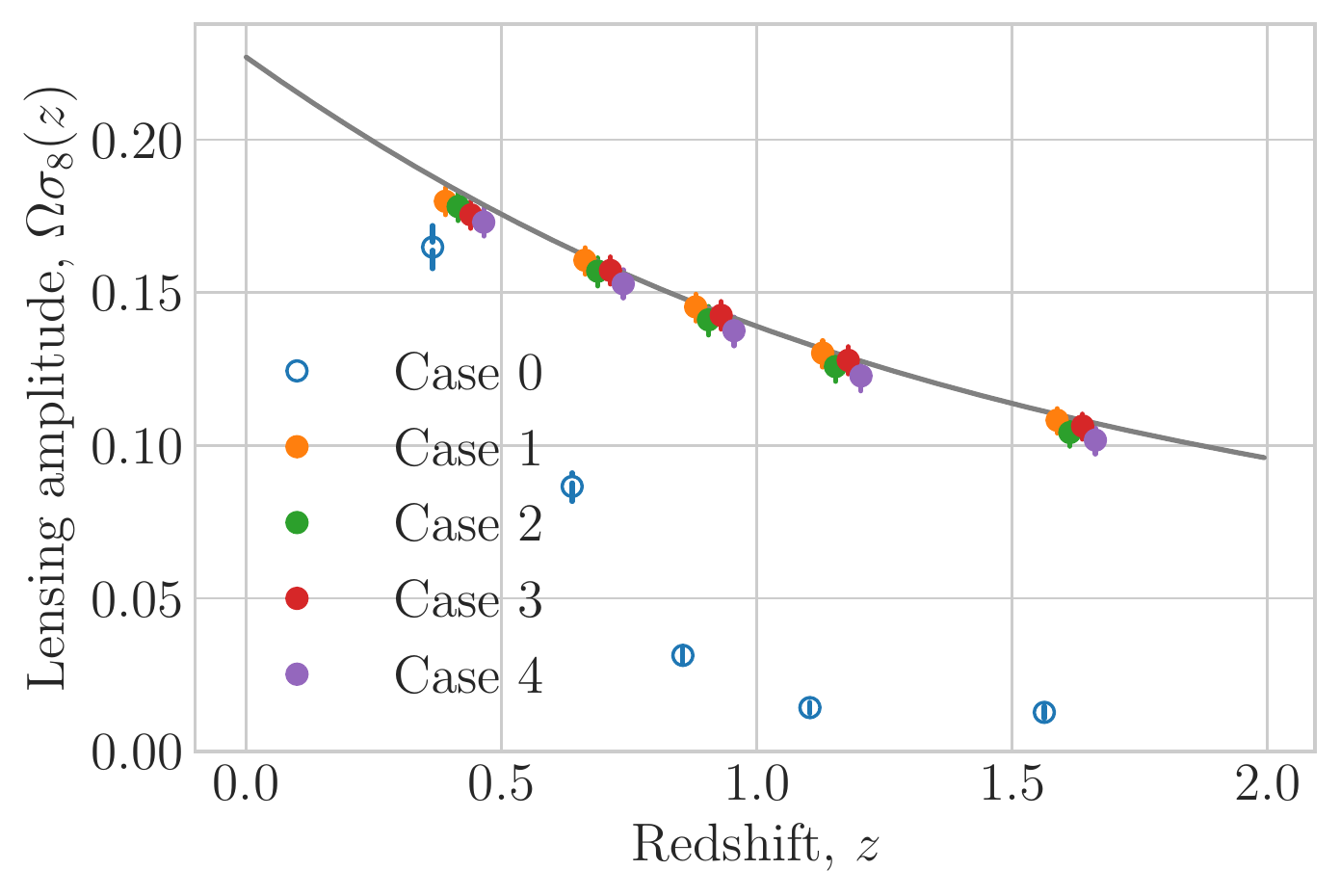}
    \caption{Input (grey curve) and reconstructed (circles with error bars) lensing amplitude \ss, for the various cases considered in this work. Data points are slightly displaced to enhance readability, with the correct \(\bar z_i\) being the central one.}
    \label{fig:s8_z}
\end{figure}

\autoref{fig:s8_z} illustrates the effectiveness of this new method. The lensing amplitude \(\ss(z)\) corresponding to the input cosmology is given by the grey curve, with reconstructed values corresponding to the coloured circles with error bars. The first thing to notice is clearly the unsuccessful recovery for Case 0. This explicitly shows how, if BNT is not performed, the strong overlap of the lensing kernels makes the factorisation of \ss\ outside of the integral no longer justified. Instead, if BNT transform is applied, I am able to reconstruct the correct value of \(\ss(z)\), even for a wrong reference cosmology, thanks to the Alcock-Paczy\'nski parameter \(\ap\).

\begin{figure*}
    \centering
    \includegraphics[width=\textwidth]{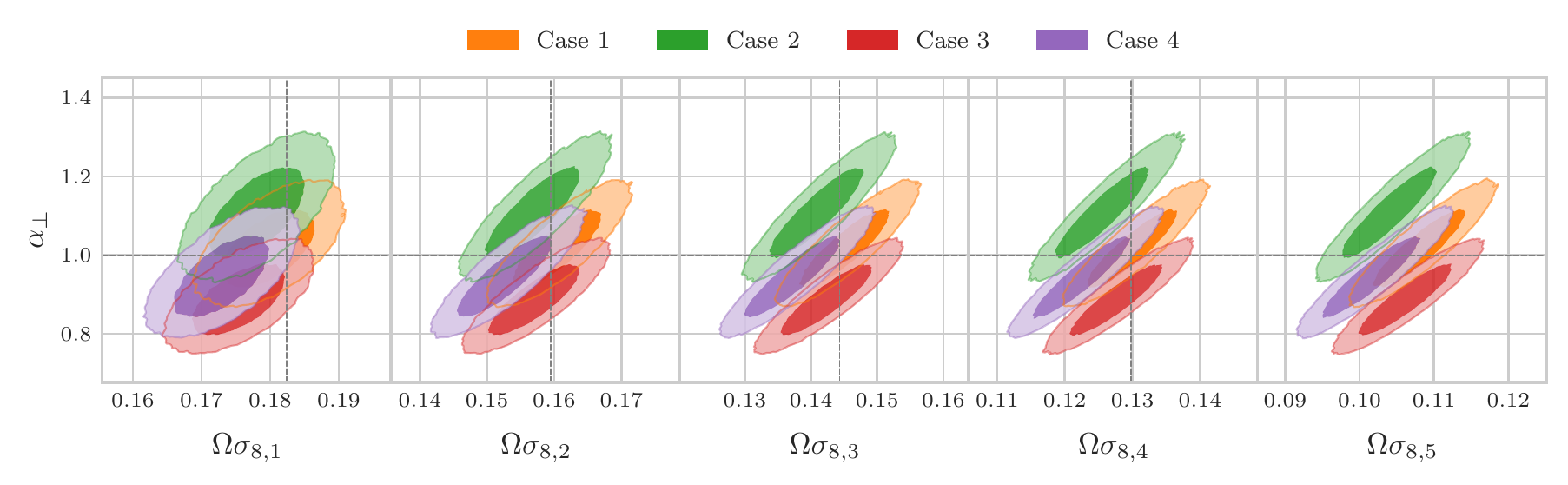}
    \caption{Two-dimensional joint marginal \(68\%\) C.L.\ contours between the redshift-binned lensing amplitudes, \(\ss[,i]\), and the transverse Alcock-Paczy\'nski parameter, \(\ap\). Horizontal and vertical dashed grey lines mark the input values, whilst colour refer to the different reference cosmologies listed in \autoref{tab:params}. Note that Case 0, namely fiducial reference cosmology with BNT transform applied to neither data nor template, is not shown due to the bad recovery of the signal.}
    \label{fig:alpha_plot}
\end{figure*}

In fact, \(\ap\) is a crucial ingredient of our template fit. \autoref{fig:alpha_plot} shows the joint marginal \(68\%\) and \(95\%\) confidence level (C.L.) contours on the \(\ss[,i]\)'s and \(\ap\). The vertical dashed grey lines mark the correct values of the \ss\ in each bin, whilst the horizontal one corresponding to \(\ap=1\) is there just to guide the eye, marking the fiducial cosmology. From this plot, we can see how the lensing amplitude is always correctly reconstructed, thanks to \ap\ accounting for any possible mismatch between the real cosmology and the one used as reference to compute the template \(\tens T^{\epsilon\epsilon}\). As expected, the recovered \ap\ is consistent with unity when the reference cosmology used for the template fit corresponds to the real one (orange contours). Note that in this figure I decided not to show contours for Case 0, as their inclusion implies an increase in axis range that would make the other cases hard to visualise.

\begin{figure}
    \centering
    \includegraphics[width=\columnwidth]{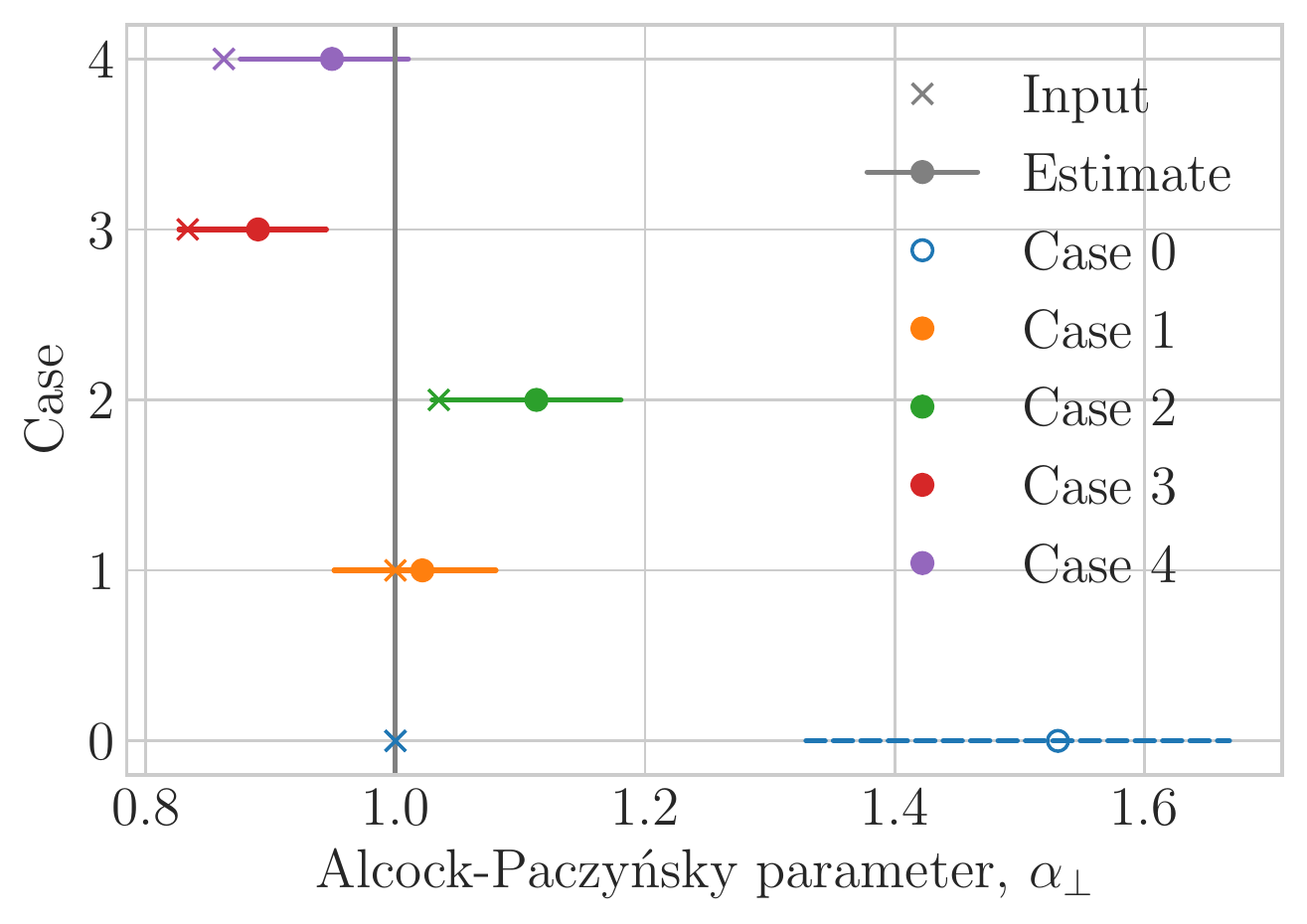}
    \caption{Input and estimates of the \ap\ parameter in the various cases.}
    \label{fig:alpha_plot_2}
\end{figure}

To appreciate this better, in \autoref{fig:alpha_plot_2} I show the input and recovered values of \(\ap\) (cross and circle markers with error bars, respectively) for the various cases listed in \autoref{tab:params}, including Case 0. Technically, \(\ap\) is a redshift-dependent quantity, in that it contains the angular diameter distance, \(D_{\rm A}(z)\). For this reason, to enable a sensible comparison of its value reconstructed from the chi-squared minimisation against the reference cosmology in input, I compute and show as crosses the average \(\langle\ap(\bar z_i)\rangle\). The vertical, solid grey line marks the scenario where the reference cosmology corresponds to the real Universe, in which case \(\ap=1\). Clearly, the agreement is excellent for all cases but Case 0.

Finally, let us scrutinise the dependency of the results upon some of the assumptions of our analysis:
\begin{itemize}
    \item \(\ell_{\rm max}\): In analysing real observations, one could implement the redshift-dependent \(\ell_{\rm max}^i\) in the chi-squared by masking the data and model vectors so that they are set to zero for the \(i\)-\(j\) bin pair when \(\ell>\min_\ell\{\ell_{\rm max}^i,\ell_{\rm max}^j\}\), as done in \citet{2021arXiv210700026T}. Here, I prefer to bracket the constraining power of the method between the worst- and best-case scenarios. The former is what I have shown so far, i.e.\ the multipole range is limited to a maximum set by the smallest of the redshift-dependent \(\ell_{\rm max}^i\)'s, \(156\). Now, for the best-case scenario, I allow the chi-square to be minimised over a data set ranging up to the largest \(\ell_{\rm max}^i\), \(705\).
    I find that error bars on measurements of \(\ss(z)\) shrink by \(55\%\) to \(73\%\).
    \item \(\bar n_{\rm g}\): Hitherto, I have employed the value \(\bar n_{\rm g}=30\,\mathrm{arcmin}^{-2}\), which is a typical sample density expected for the oncoming generation of cosmic shear surveys \citep[see e.g.][]{Blanchard:2019oqi,2019ApJS..242....2C}. However, it is worth noting that the method presented here focusses on linear scales, where shot noise is expected to be largely subdominant. Therefore, I expect our results to be not strongly influenced by \(\bar n_{\rm g}\). To test this, I rerun our benchmark analysis for two more pessimistic values, \(15\,\mathrm{arcmin}^{-2}\) and \(3\,\mathrm{arcmin}^{-2}\). As envisaged, even with source samples \(50\%\) to \(90\%\) sparser, constraints on the various \(\ss[,i]\) only loosen in the range \([7,11]\%\) in the former case, and in the range \([29,56]\%\) in the latter.
    \item \(N_z\): The ultimate aim of this method, like the corresponding one for galaxy clustering, is to probe the redshift evolution of cosmic structures. Harmonic-space power spectra have the advantage to allow for different binning strategies. Hence, I now rerun the benchmark analysis with two different binnings, namely \(N_z=3\) and \(15\). I find that the variance on the average \(\ss[,i]\) scales sub-linearly with \(N_z\). This is because, the more the bins, the more the cross-bin spectra, which contain additional information---albeit partly correlated. This leaves us ample room to tailor the best course of action in any given scenario, e.g.\ opting for more bins if we look for a smooth variation of \(D(z)\) compared to \lcdm, or to choose fewer bins if we search for a specific feature at a certain redshift.
\end{itemize}

\section{Conclusions and future work.}
In this paper, I have introduced a new method to measure the growth of cosmic structures as a function of redshift directly from cosmic shear data. This is achieved with a template-fitting approach, where the redshift-dependent lensing amplitude \(D(z)\,\sigma_8\,\om\eqqcolon\ss(z)\) is factorised out of the integral that gives the tomographic shear power spectrum. For this, I implement the BNT nulling technique \citep{2014MNRAS.445.1526B}, to localise the lensing signal in redshift. Thus, I overcome the issue of the overlap of
shear kernels---even for widely separated redshift bins---due to lensing being an integrated effect.

I have demonstrated the performance of the method by running a synthetic-data analysis, where I fit the template against the full theoretical prediction for the cosmic shear power spectrum. The amplitudes \ss[,i]\ are a free parameter in each bin. On top of it, I have also added an Alcock-Paczy\'nski parameter \ap\ to account for an incorrect choice of reference cosmology in the computation of the template. This method performs very well, being able to sample the underlying curve \(\ss(z)\) with a precision of a few per cent in all the bins that the source redshift distribution has been sliced into.

To anybody familiar with cosmic shear analyses, there is an elephant in the room here, and that is intrinsic alignments (IA). IA are a contaminant to the cosmological shear signal, due to the fact that galaxies do preferentially align with each other along their major axes/angular momentum directions due to interactions with the tidal gravitational field in the local environment \citep[see][for thorough reviews on the topic]{2015SSRv..193....1J,2015SSRv..193...67K,2015SSRv..193..139K}. As a result, the observed galaxy ellipticities are a combination of the underlying cosmological shear and IA. As such, the measured power spectrum contains additional terms on top of those in \autoref{eq:Cl}.

Considering that modelling IA is a well-known problem in the analysis of cosmic shear surveys, I argue that the template-fitting approach described here can be readily modified to encapsulate IA. This could be done by promoting IA redshift-dependent amplitudes to additional nuisance parameters, with the inclusion of the \ap\ rescaling to the IA power spectra as well. In a sense, this is not dissimilar from what happens in the case of galaxy clustering, where the dominant term due to Newtonian density fluctuations is complemented by additional contributions, such as redshift-space distortions and lensing magnification \citep[see][]{2021arXiv210700026T,Camera:2022}. Such an approach will be scrutinised in a follow-up paper.

\begin{acknowledgments}
The author warmly thanks Ruth Durrer, Gigi Guzzo, Tom Kitching, and Roy Maartens for their valuable comments on the manuscript, as well as Matteo Martinelli for black-belt tips on the usage of \texttt{GetDist}. The author acknowledges support from the `Departments of Excellence 2018-2022' Grant (L.\ 232/2016) awarded by the Italian Ministry of University and Research (\textsc{mur}).
\end{acknowledgments}


\bibliography{biblio}

\end{document}